# Network Compression: Memory-Assisted Universal Coding of Sources with Correlated Parameters

Ahmad Beirami and Faramarz Fekri
School of Electrical and Computer Engineering
Georgia Institute of Technology, Atlanta GA 30332, USA
Email: {beirami, fekri}@ece.gatech.edu

*Abstract*—The existence of significant amount of correlation in the network traffic has stimulated the development of in-network traffic reduction techniques since end-to-end universal compression solutions would not perform well over Internet packets due to the finite-length nature of data. Recently, we proposed a memory-assisted universal compression technique that holds a significant promise for reducing the amount of traffic in the networks. The idea is based on the observation that if a finite-length sequence from a server (source) is to be compressed and transmitted over the network, the associated universal code entails a substantial overhead. On the other hand, intermediate nodes can reduce the transmission overhead by *memorizing* the source statistics when forwarding the sequences from the previous communications with the server. In this paper, we extend this idea to the scenario where multiple servers are present in the network by proposing *distributed network compression via memory*. We consider two spatially separated sources with correlated unknown source parameters. We wish to study the universal compression of a sequence of length $n$ from one of the sources provided that the decoder has access to (i.e., memorized) a sequence of length $m$ from the other source. In this setup, the correlation does not arise from symbol-by-symbol dependency of two outputs from the two sources (as in Slepian-Wolf setup). Instead, the two sequences are correlated because they are originated from the two sources with *unknown* correlated parameters. The finite-length nature of the compression problem at hand requires considering a notion of almost lossless source coding, where coding incurs an error probability $p_e(n)$ that vanishes as sequence length $n$ grows to infinity. We obtain bounds on the redundancy of almost lossless codes when the decoder has access to a random memory of length $m$ as a function of the sequence length $n$ and the permissible error probability $p_e(n)$. Our results demonstrate that distributed network compression via memory has the potential to significantly improve over conventional end-to-end compression when sufficiently large memory from previous communications is available to the decoder.

## I. INTRODUCTION

Several networking applications involve acquiring data from multiple distributed (i.e., spatially separated) sources that cannot communicate with each other. These applications include acquiring digital/analog data from sensors [1]–[5], the CEO problem [6], [7], delivery of network packets in a content-centric network [8], acquiring data from femtocell wireless networks [9], [10], acquiring data chunks from the cloud [11], [12], etc. What is perhaps common in all of the above problems is the *bandwidth limitation*, i.e., there is a fundamental capacity for the information that can be transmitted in the network infrastructure. Hence, data compression can significantly improve the performance in any of such applications.

The premise of data compression broadly relies on the data being correlated. As one example, when data is gathered from multiple sensors that measure the same phenomenon (e.g., temperature), the readings from the sensors are clearly correlated. As another example, when chunks of the same file/content are acquired by a client in a content-centric network, the data chunks are correlated as they are originated from the same data server. Further, data that is originated from a *mirror server* is correlated with data that comes from the original server. The focus of this work is on the reduction of the wireless/wired data traffic from multiple sources by utilizing such correlations. The scope of this work is significant as high correlation levels as much as 90% have been reported in the wired/wireless Internet traffic data [13]–[15], which has motivated a lot of research so as to reduce the traffic by utilizing such correlations.

Existing solutions that utilize such correlations in order to reduce the data transmission in the Internet are limited in scope. Application-level content caching [16] cannot utilize the packet-level redundancy and statistical correlations across the contents. Packet-level redundancy elimination techniques [17] are ad-hoc in nature and can only remove duplicates of a big chunk of the data packet while they ignore the statistical correlations in the packet-level. Application-level universal compression [18]–[20] techniques do not utilize packet-level redundancies and more importantly cannot utilize the correlations in data that are originated from spatially separated sources. Packet-level memory-assisted compression techniques [14], [21], [22] utilize the statistical correlation among the packets while its extension to multiple sources is not readily available.

In this paper, we introduce and study *distributed network compression via memory*, where we assume that the unknown parameter vectors of the distributed sources follow a correlated statistical model. By distributed we mean that the sources are spatially separated and the encoders do not communicate with each other. We stress that the nature of our problems in network compression involving multiple sources is fundamentally different from those addressed by the Slepian-Wolf (SW) coding and multi-terminal source coding in [1], [3], [6], [7]. Here, instead of symbol-by-symbol correlation between the sequences as in SW setup or the correlated Gaussian model among several observations of a phenomenon, the correlation is due to the the source parameters being a priori unknown [23], [24]. To clarify, considering the example in Fig. 1 with sources $S_1$ and $S_2$, $y^m$ and $x^n$ would be independent given that the source models are known. However, when the source parameter is unknown, $y^m$ and $x^n$ are *correlated* with each other through the information they contain about the unknown but correlated source parameters. The question, which incurs in distributed network compression via memory, is whether or not this correlation can be potentially leveraged by the encoder of $S_2$ and the decoder at $M$ in the decoding of $x^n$ using $y^m$ (from $S_1$) to reduce the codelength of $x^n$.

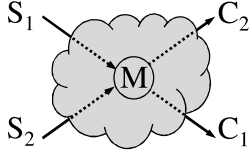

Fig. 1. The basic scenario of two-source memory-assisted compression.

The rest of this paper is organized as follows. In Section II, we present the problem setup and the related work. In Section III, we briefly review the necessary background and definitions. In Section IV, we present our main results on the redundancy. In Section V, we provide discussion on the results. Finally Section VI concludes the paper.

## II. PROBLEM SETUP AND RELATED WORK

We present the memory-assisted network compression problem in the most basic scenario, shown in Fig. 1, consisting of two correlated sources located in nodes $S_1$ and $S_2$, the intermediate relay node $M$, and two client nodes $C_1$ and $C_2$. Let $y^m$ and $x^n$ denote two sequences with lengths $m$ and $n$ that are generated by $S_1$ and $S_2$, respectively. We assume that $S_1$ has transmitted the sequence $y^m$ to $C_1$ through the intermediate node $M$. We further assume that $M$ is a memory unit, i.e., capable of memorizing the sequence $y^m$. Next, at some later time, $S_2$ wishes to send $x^n$ to $C_2$ through the intermediate node $M$. At this time, $y^m$ is available to the decoder at $M$. Thus, the encoder at $S_2$ can encode the sequence $x^n$ with the knowledge that $y^m$ is available to the decoder at $M$, potentially improving the universal compression of $x^n$ on the path from $S_2$ to $M$. Such a code is decoded by $M$ before being forwarded to the final destination $C_2$. A trivial lower bound on the expected number of bits necessary for transmitting $x^n$ on the $S_2$-$M$ path will be $H(X^n|Y^m)$. Our goal is to analyze the lower bound and its achievability in various settings.

Slepian and Wolf already demonstrated that if the data streams from two sources $S_1$ and $S_2$ have symbol-by-symbol correlation, the sequences can be compressed to their joint entropy when decoded at $M$ [1]. The idea is based on compressing the jointly typical sequences $(x^n, y^n)$. As the length $n$ of the sequences increases to infinity, the decoding of the sequence $x^n$ at $M$ can be performed using an almost lossless code with the average length that asymptotically approaches the conditional entropy, i.e., $H(X^n|Y^n)$, with asymptotically zero error probability, i.e., $\lim_{n\to\infty} p_e(n) = 0$. On the other hand, if the decoder at $M$ chooses not to utilize the side information provided by the sequence $y^n$ or the coding is performed strictly lossless,[1] the encoder at $S_2$ would have to encode the sequence $x^n$ irrespective to what has already been communicated between $S_1$ and $M$, which would in turn result in an average code length of $H(X^n)$. After relatively recent development of practical Slepian-Wolf (SW) coding schemes by Pradhan and Ramchandran [2], SW coding has drawn a great deal of attention as a promising compression technique in many applications such as sensor networks (cf. [4] and the references therein) and distributed video coding [5].

The Slepian-Wolf theorem naturally suits applications where the (new) sequence $x^n$ from $S_2$ (in Fig. 1) can be viewed as a noisy version of the (previously seen) sequence $y^m$, such as data gathering from neighboring sensors that measure the same phenomenon. However, in many other scenarios, the compression of spatially separated sources cannot be modeled by the SW framework. Examples include the universal compression of data from multiple mirrors of a data server and acquiring data chunks in a content-centric network. In such applications, it is plausible to assume that the sources ($S_1$ and $S_2$ in Fig. 1) follow a correlated (sometimes even identical) statistical model that is a priori unknown (to the encoder and the decoder) requiring universal compression [23], [26], [27]. We assume that the servers at $S_1$ and $S_2$ are stationary and ergodic parametric information sources that are unknown to the coding scheme. The following example clarifies this model of correlation.

As an example, assume that source $S_1$ is a server that generates Bernoulli random variables (RVs) with unknown source parameter $\theta$. Further, assume that source $S_2$ is a mirror server in a different location with very similar content. Thus, source $S_2$ is a Bernoulli RV generator with parameter $\phi$, where we assume that $\phi$ follows a Gaussian distribution around $\theta$. (If the mirror servers contain the exact same content we may even assume that $\phi = \theta$, i.e., the variance of $\phi$ can be assumed to be equal to zero). Let the sequences $y^m$ and $x^n$ be generated independently by the two servers $S_1$ and $S_2$, respectively. In this setup, the sequence $y^m$ is correlated with $x^n$ through the information that they carry about the unknown source parameters. For example, if most of the bits in $y^m$ are 1's, it is very likely that most of the bits in $x^n$ are also 1's. The question is, assuming two sources $S_1$ and $S_2$ with correlated unknown parameters and having $y^m$ from $S_1$ memorized at the decoder at $M$, what is the achievable universal compression performance on $x^n$ at $S_2$-$M$ path and whether the correlation between $x^n$ and $y^m$ can be potentially leveraged by the encoder of $S_2$ and the decoder at $M$ in the decoding of $x^n$ using $y^m$ to reduce the codelength of $x^n$.

---

[1]Please see [25] for the formal definition of strictly lossless and almost lossless codes. In short, the strictly lossless coding is more restrictive than almost lossless coding since it requires $\forall n; p_e(n) = 0$ as apposed to $\lim_{n\to\infty} p_e(n) = 0$.

This problem can also be viewed as universal compression [18]–[20] with training data that is only available to the decoder. In [21], [22], we theoretically derived the *gain* that is obtained in the universal compression of the new sequence $x^n$ from $S_2$ by memorizing (i.e., having access to) $y^m$ from $S_1$ at *both* the decoder (at $M$) and the encoder (at $S_2$). This corresponds to the reduced case of our problem where the sources $S_1$ and $S_2$ are either co-located (a single source) or allowed to communicate. For the reduced problem case, in [14], [28], we further extended the setup to a network with a single source and derived bounds on the *network-wide gain* where a small fraction of the intermediate nodes in the network are capable of memorization. However, the extension to the multiple spatially separated sources, where the training data is only available to the decoder, is non-trivial and raises a new set of challenges that we aim to address.

In [25], we extended the network compression to distributed identical sources in the special case where the sources were identical. We derived an upper bound on the achievable average minimax redundancy, where $S_1$ and $S_2$ share an indexical parameter vector. In this paper, we let the information sources at $S_1$ and $S_2$ be parametric with $d$-dimensional parameter vectors $\theta$ and $\phi$, respectively. These parameter vectors are unknown a priori to the encoder and the decoder. Throughout the paper, we refer to this problem setup as Distributed Network Compression with Correlated Parameters (DNC-CP). We stress that the nature of DNC-CP is fundamentally different from those addressed by the Slepian-Wolf (SW) theorem in [1]. Here, instead of symbol-by-symbol correlation between the sequences as in SW setup, we target to remove the redundancy incurred by the universal compression of finite-length sequences, whose dependency is due to the correlation of their unknown source parameters that are a priori unknown [21], [23], [24]. Note that as the length of the sequence $x^n$ grows to infinity, the redundancy rate in the compression of $x^n$ vanishes since $\frac{1}{n} H(X^n)$ converges to the entropy rate as $n \to \infty$, and hence, the potential benefits of DNC-CP vanish as the sequence length grows, which contrasts the Slepian-Wolf framework where the benefits are studied in the asymptotic regime.

## III. NOTATIONS AND DEFINITIONS

Thus far, we described the basic problem setup. In this section, we provide further details involving notations and definitions. Following the notation in [25], let $\mathcal{A}$ be a finite alphabet. Let $d$ be the number of the source parameters. Let $\Theta^d$ denote the space of $d$-dimensional parameter vectors. Let $\lambda \in \Theta^d$ denote a $d$-dimensional parameter vector. Let $\mathcal{P}^d$ denote the *family* of sources that can be described with a $d$-dimensional unknown parameter vector $\lambda$. We denote $\mu_\lambda$ as the probability measure that is defined by the parameter vector $\lambda$ under the parametric source model. Let $\mathcal{I}(\lambda)$ denote the Fisher information matrix for parameter vector $\lambda$.

We assume that the parameter vector $\theta \in \Theta^d$ (corresponding to source $S_1$) follows the worst-case prior in the sense that it maximizes the expected redundancy (i.e., the capacity achieving prior in the maximin sense). This prior distribution is particularly interesting because it corresponds to the worst-case compression performance for the best compression scheme. We further assume that given $\theta$, the parameter vector $\phi \in \Theta^d$ (i.e., the parameter vector of source $S_2$) follows a Gaussian distribution with mean $\theta$ and covariance matrix $\Gamma(\theta)$. This models the nature of the correlation of the sources $S_1$ and $S_2$ in our setup. Let $\mathcal{J}(\theta)$ be a $d \times d$ matrix associated with the parameter vectors $\phi$ and $\theta$, defined as $\mathcal{J}(\theta) \triangleq \Gamma(\theta) \mathcal{I}(\theta)$. We assume that $\mathcal{J}(\theta)$ is a positive definite matrix. This assumption is necessary for the conditional distribution to be well defined. Let $I_d$ be the $d \times d$ identity matrix. We use the notation $x^n = (x_1, ..., x_n) \in \mathcal{A}^n$ to present a sequence of length $n$ from the alphabet $\mathcal{A}$ generated by $S_2$. We further denote $X^n$ as a random sequence of length $n$ that follows the probability distribution $\mu_\phi$. Let $H_n(\phi)$ be the entropy of the source $S_2$ given the parameter vector $\phi$, i.e., $H_n(\phi) = H(X^n|\phi) = \mathbf{E} \log \left( \frac{1}{\mu_\phi(X^n)} \right)$.[2]

Let $c_n : \mathcal{A}^n \to \{0,1\}^*$ be an injective mapping from the set $\mathcal{A}^n$ of the sequences of length $n$ over $\mathcal{A}$ to the set $\{0,1\}^*$ of binary sequences. Further, let $l_n^{p_e}(x^n)$ denote the almost lossless length function of the codeword associated with the sequence $x^n$ with permissible error $p_e$. In the study of coding strategies for DNC-CP, we compare the following relevant cases for the compression of the sequence $x^n$ from $S_2$ provided that the sequence $y^m$ from $S_1$ has already been memorized by the node $M$ (in Fig. 1).

- UComp (Universal compression without memorization), which only applies lossless universal compression to $x^n$ at $S_2$ without using the side information $y^m$ at $M$.
- DUCompMD (Distributed universal compression with memory at decoder), which assumes that decoder (at $M$) has access to context memory sequence $y^m$ while the encoder (at $S_2$) only knows $m$ but does not know the exact sequence $y^m$. The encoder then applies a universal code to $x^n$ that is decoded at $M$ by utilizing $y^m$.
- DUCompME (Distributed universal compression with common memory at both the decoder and the encoder), which assumes that the two encoders at $S_1$ and $S_2$ can communicate, and thus, the decoder (at $M$) and the encoder (at $S_2$) have access to a shared sequence $y^m$, which is utilized in the compression of $x^n$ at $S_2$.

In this paper, we use the average minimax redundancy as the performance metric for the different coding strategies. Let $L_n^{p_e}$ denote the space of universal almost lossless length functions on a sequence of length $n$, with permissible decoding error $p_e$. Denote $R_n(l_n^{p_e}, \phi)$ as the expected redundancy of the almost lossless code on a sequence of length $n$ for the parameter vector $\phi$, i.e., $R_n(l_n^{p_e}, \phi) = \mathbf{E}l_n^{p_e}(X^n) - H_n(\phi)$. Accordingly, the average minimax redundancy, which corresponds to the performance of the best code over the worst parameter vector is defined as follows.

---
[2]Throughout this paper, all expectations are taken with respect to the probability measure $\mu_\phi$, and $\log(\cdot)$ denotes the logarithm in base 2.

$$\bar{R}_{\text{UComp}}^{p_e}(n) \triangleq \inf_{l_n^{p_e} \in L_n^{p_e}} \sup_{\theta \in \Theta^d} R_n(l_n^{p_e}, \theta). \quad (1)$$

We denote $\bar{R}_{\text{UComp}}^0(n)$ as the average minimax redundancy when the compression scheme is restricted to be strictly lossless instead of almost lossless, i.e., $p_e = 0$.

In DUCompMD, let $\hat{l}_{n,m,\Gamma}^{p_e} : \mathcal{A}^n \times \mathbb{N} \times \mathbb{R}^{d \times d} \to \mathbb{R}$. Note that in this case, the sequence $y^m$ is not known to the encoder while the length $m$ is still available to the encoder. Denote the lossless universal length function with a memorized sequence of length $m$ that is only available to the decoder with permissible error probability $p_e$. Further, denote $\hat{L}_{n,m,\Gamma}^{p_e}$ as the space of such lossless universal length functions. Denote $R_n(\hat{l}_{n,m,\Gamma}^{p_e}, \theta)$ as the expected redundancy of encoding a sequence $x^n$ of length $n$ using the length function $\hat{l}_{n,m,\Gamma}^{p_e}$. Further, let $\bar{R}_{\text{DUCompMD}}^{p_e}(n, m, \Gamma)$ denote the expected minimax redundancy, i.e.,

$$\bar{R}_{\text{DUCompMD}}^{p_e}(n, m, \Gamma) \triangleq \inf_{\hat{l}_{n,m,\Gamma}^{p_e} \in \hat{L}_{n,m,\Gamma}^{p_e}} \sup_{\theta \in \Theta^d} R_n(\hat{l}_{n,m,\Gamma}^{p_e}, \theta). \quad (2)$$

Likewise, let $l_{n,m,\Gamma}^{p_e} : \mathcal{A}^n \times \mathcal{A}^m \times \mathbb{R}^{d \times d} \to \mathbb{R}$ be the lossless universal length function with a shared memory of length $m$ and permissible error probability $p_e$ and covariance matrix $\Gamma$. Denote $L_{n,m,\Gamma}^{p_e}$ as the space of lossless universal length functions on a sequence of length $n$ with a shared memory of length $m$. Denote $R_n(l_{n,m,\Gamma}^{p_e}, \theta)$ as the expected redundancy of encoding a sequence of length $n$ form the source using the length function $l_{n,m,\Gamma}^{p_e}$. Let $\bar{R}_{\text{DUCompME}}^{p_e}(n, m, \Gamma)$ denote the expected minimax redundancy for the lossless universal length function with a memory size of length $m$ shared between the encoder and the decoder, i.e.,

$$\bar{R}_{\text{DUCompME}}^{p_e}(n, m, \Gamma) \triangleq \inf_{l_{n,m,\Gamma}^{p_e} \in L_{n,m,\Gamma}^{p_e}} \sup_{\theta \in \Theta^d} R_n(l_{n,m,\Gamma}^{p_e}, \theta). \quad (3)$$

Again, when we set $p_e = 0$ we refer to the strictly lossless case. The following is a trivial statement comparing the performance of almost lossless coding versus strictly lossless coding.

**Fact 1** *For all of of the described coding strategies, the strictly lossless redundancy is an upper bound on the the redundancy of the almost lossless coding for any $p_e$.*

The following trivial inequalities demonstrate that the redundancy decreases when side information is available to the decoder. Moreover, if the side information is also available to the decoder, the redundancy is further decreased.

**Fact 2** *Let $p_e \geq 0$. Then, we have*
$$\bar{R}_{DUCompME}^{p_e}(n, m, \Gamma) \leq \bar{R}_{DUCompMD}^{p_e}(n, m, \Gamma) \leq \bar{R}_{UComp}^{p_e}(n).$$

## IV. Main Results

In this section, we evaluate the performance of each of the different coding schemes introduced in the previous section for the DNC-CP problem using their corresponding average minimax redundancy for both almost lossless and strictly lossless codes. We treat the strictly lossless codes (i.e., $p_e = 0$) separately since they are interesting on their own. Some of the proofs are omitted due to the lack of space. All these results are valid for finite-length $n$ (as long as $n$ is large enough to satisfy the central limit theorem criteria).

### A. Strictly Lossless DNC-CP

*1) UComp:* In this case, the side information sequence is not utilized at the decoder for the compression of $x^n$, and hence, the minimum number of bits required to represent $x^n$ is $H(X^n) = H(X^n|\phi) + I(X^n; \phi)$. Thus, $\bar{R}_{\text{UComp}}^0(n) = \sup_{\omega(\phi)} I(X^n; \phi)$. Thus, it is straightforward to show the following [24], [29]

**Theorem 1** *The average minimax redundancy for strictly lossless UComp coding strategy is*

$$\bar{R}_{\text{UComp}}^0(n) = \frac{d}{2} \log\left(\frac{n}{2\pi e}\right) + \log \int_{\phi \in \Theta^d} |\mathcal{I}(\phi)|^{\frac{1}{2}} d\phi + O\left(\frac{1}{n}\right).$$

*2) DUCompMD:* Next, we confine ourselves to strictly lossless codes in the DUCompMD strategy. In [25], we established a result that the memorization of $y^m$ at the decoder does not provide any benefit on the strictly lossless universal compression of the sequence $x^n$ from $S_2$ when the parameter vectors are identical. It is straightforward to generalize that result as the following.

**Theorem 2** *The average minimax redundancy for strictly lossless DUCompMD coding strategy is*

$$\bar{R}_{DUCompMD}^0(n, m, \Gamma) = \bar{R}_{UComp}^0(n).$$

*3) DUCompME:* Next, we present the main result on the strictly lossless codes for DUCompME coding strategy. In this case, since a random sequence $Y^m$ is also known to the encoder, the achievable codelength for representing $x^n$ is given by $H(X^n|Y^m)$. Then, the redundancy is given by the following theorem.

**Theorem 3** *The average minimax redundancy for strictly lossless DUCompME coding strategy is*

$$\bar{R}_{DUCompME}^0(n, m, \Gamma) = \hat{R}(n, m, \Gamma) + O\left(\frac{1}{n} + \frac{1}{m}\right),$$

*where the main redundancy term is given by*

$$\hat{R}(n, m, \Gamma) = \sup_\phi \frac{1}{2} \log\left|\left(1 + \frac{n}{m}\right) I_d + n\mathcal{J}(\phi)\right|. \quad (4)$$

## B. Almost Lossless DNC-CP

In this case, we investigate the reduction in the average codelength associated with a sequence $x^n$ as a result of the permissible error probability $p_e$.

*1) UComp:* We demonstrate the following lower bound on the redundancy.

**Theorem 4** *The average minimax redundancy for almost lossless UComp coding strategy is lower bounded by*

$$\bar{R}^{p_e}_{UComp}(n) \geq (1-p_e)\bar{R}^{0}_{UComp}(n) - h(p_e) - p_e H_n(\phi).$$

*Proof:* Please refer to the Appendix for the proof. ■

*2) DUCompMD:* In this case, we proved in [25] that the permissible error probability $p_e$ *potentially* results in further reduction in the average codelength. The generalization of that result for the sources with correlated parameters is given by the following theorem.

**Theorem 5** *The average minimax redundancy for almost lossless DUCompMD coding strategy is upper bounded by*

$$\bar{R}^{p_e}_{DUCompMD}(n,m,\Gamma) \leq \hat{R}(n,m,\Gamma)+\mathcal{F}(d,p_e)+O\left(\frac{1}{m}+\frac{1}{n}\right),$$

*where $\hat{R}(n,m,\Gamma)$ is the main redundancy term defined in (4) and $\mathcal{F}(d,p_e)$ is the penalty due to the encoders not communicating given by*

$$\mathcal{F}(d,p_e) = \frac{d}{2}\log\left(1+\frac{2}{d\log e}\log\frac{1}{p_e}\right). \quad (5)$$

*3) DUCompME:* We have the following lower bound.

**Theorem 6** *The average minimax redundancy for almost lossless DUCompME coding strategy is upper bounded by*

$$\bar{R}^{p_e}_{DUCompME}(n,m,\Gamma) \geq (1-p_e)\bar{R}^{0}_{DUCompME}(n,m,\Gamma)$$
$$- h(p_e) - p_e H_n(\phi).$$

## V. DISCUSSION ON THE RESULTS

In this section, we provide some discussion on the significance of the results for different DNC-CP coding strategies. We discuss the strictly lossless case followed by two examples that illustrate the impact of the source parameter correlation on the results of the almost lossless and strictly lossless schemes.

## A. Strictly Lossless

In the case of UComp, Theorem 1 determines the achievable average minimax redundancy for the compression of a sequence of length $n$ encoded regardless of the previous sequence $y^m$. In other words, UComp is an end-to-end universal compression scheme which does not use memorization. Hence, UComp is used as the benchmark for the performance of DUCompMD and DUCompME, which are memory-assisted network compression techniques.

According to Theorem 2, in DNC-CP, if strictly lossless codes are to be used for the compression of $x^n$ from $S_2$, the memorization of the previous sequences from $S_1$ by the decoder does not provide any benefit, assuming that the two encoders at $S_1$ and $S_2$ do not communicate (i.e., DUCompMD). In other words, the best that $S_2$ can do for the strictly lossless compression of $x^n$ is to simply apply a traditional universal compression.

Theorem 3 determines the main redundancy term in the strictly lossless DUCompME coding strategy. It can be deduced from Fact 2 thatthat if the two encoders communicate (i.e., DUCompME), the performance of strictly lossless compression of $x^n$ would improve with respect to UComp. It is straightforward to see that as $m$ grows, the main redundancy term in (4) decreases. However, the main redundancy term for very large memory (i.e., $m \to \infty$) is given by

$$\hat{R}(n,\infty,\Gamma) = \sup_\lambda \frac{1}{2}\log|I_d + n\mathcal{J}(\lambda)|, \quad (6)$$

which remains non-zero in general. Therefore, increasing $m$ beyond a certain limit does not provide further performance improvement. In summary, for the strictly lossless case, only DUCompME is interesting as it offers benefit over UComp but it is not practical as it requires the encoders to communicate.

## B. Example 1: Identical Source Parameters

In this special case, we assume that the source parameters $\theta$ and $\phi$ are identical, and hence, $\mathcal{J}(\theta) = \Gamma(\theta) = \mathbf{0}_d$. The performance of strictly lossless DUCompME coding strategy and the almost lossless DUCompMD coding strategy is quantified by $\hat{R}(n,m,\mathbf{0}_d)$, which is given in the following proposition, giving back what was proved in [25].

**Proposition 7** *The main redundancy term of (4) for the identical source parameters is given by*

$$\hat{R}(n,m,\mathbf{0}_d) = \frac{d}{2}\log\left(1+\frac{n}{m}\right).$$

We further consider the redundancy for large $m$. It can be shown that we have $\lim_{m\to\infty}\bar{R}^0_{DUCompME}(n,m,\mathbf{0}_d) = 0$. In other words, since the parameter vector will be known to both the encoder and the decoder, the code's redundancy vanishes similar to the Shannon code.[3] In this case, the fundamental limits are those of known source parameters and universality no longer imposes a compression overhead.

---
[3]Note that we have ignored the integer constraint on the length functions in this paper, which will result in a negligible $O(1)$ redundancy that is exactly analyzed in [30], [31].

*C. Example 2: Correlation Covariance Matrix Inversely Proportional to Fisher Information Matrix*

Next, we consider the case where the covariance matrix $\Gamma(\theta)$ is inversely proportional to the Fisher information matrix, i.e., $\Gamma^{-1}(\theta) = \alpha \mathcal{I}(\theta)$. In this case, the two parameter vectors can be viewed as estimates of each other.

**Proposition 8** *The main redundancy term of (4) for the case where $\Gamma^{-1}(\theta) = \alpha \mathcal{I}(\theta)$ is given by*

$$\hat{R}(n, m, \frac{1}{\alpha}\mathcal{I}^{-1}) = \frac{d}{2}\log\left(1 + \frac{n}{m} + \frac{n}{\alpha}\right).$$

Hence, as the correlation between the two parameters increases, the redundancy decreases and eventually converges to that of the identical source parameters.

## VI. CONCLUSION

In this paper, we introduced and studied the problem of Universal Compression of Distributed Sources with Correlated Parameters (DNC-CP). In DNC-CP, the correlation of the two source parameters becomes relevant due to the finite-length universal compression constraint. This model departs from the nature of the correlation in the SW framework. For DNC-CP, involving two correlated sources, we investigated the average minimax redundancy. We demonstrated that memorization at the intermediate nodes in the network can help to noticeably improve the performance of the universal compression on multiple sources whose parameters are correlated. On the other hand, we did not provide a coding strategy that achieves the performance limits derived in this paper.

## APPENDIX

## PROOF OF THEOREM 4

In order to prove this theorem, we consider $H(X^n, \hat{X}^n, \mathbf{1}_e(X^n))$. Note that both $\hat{X}^n$ and $\mathbf{1}_e(X^n)$ are deterministic functions of $X^n$ and hence

$$H(X^n, \hat{X}^n, \mathbf{1}_e(X^n)) = H(X^n). \quad (7)$$

On the other hand, we can also use the chain rule in a different order to arrive at the following.

$$H(X^n, \hat{X}^n, \mathbf{1}_e(X^n)) = H(\hat{X}^n) + H(\mathbf{1}_e(X^n)|\hat{X}^n) \\ + H(X^n|\mathbf{1}(X^n), \hat{X}^n). \quad (8)$$

Hence,

$$H(\hat{X}^n) = H(X^n) - H(\mathbf{1}_e(X^n)|\hat{X}^n) - H(X^n|\mathbf{1}(X^n), \hat{X}^n)$$
$$\geq H(X^n) - h(p_e) - H(X^n|\mathbf{1}(X^n), \hat{X}^n) \quad (9)$$
$$\geq H(X^n) - h(p_e) - p_e H(X^n), \quad (10)$$

where the inequality in (9) is due to the fact that $H(\mathbf{1}_e(X^n)|\hat{X}^n) \leq H(\mathbf{1}_e(X^n)) = h(p_e)$ and the inequality in (10) is due to Lemma 1.

**Lemma 1** $H(X^n|\mathbf{1}_e(X^n), \hat{X}^n) \leq p_e H(X^n).$

*Proof:*

$$H(X^n|\mathbf{1}_e(X^n), \hat{X}^n) = (1 - p_e)H(X^n|\mathbf{1}_e(X^n,) = 0, \hat{X}^n) \\ + p_e H(X^n|\mathbf{1}_e(X^n) = 1, \hat{X}^n) \quad (11)$$
$$\leq p_e H(X^n). \quad (12)$$

The first term in (11) is zero since if $\mathbf{1}_e(X^n) = 0$, we have $X^n = \hat{X}^n$ and hence $H(X^n|\mathbf{1}_e(X^n,) = 0, \hat{X}^n) = 0$. The inequality in (12) then follows from the fact that $H(X^n|\mathbf{1}_e(X^n) = 1, \hat{X}^n) \leq H(X^n)$ completing the proof. ∎

The proof of the theorem is completed by noting that $H(X^n) = H_n(\theta) + \bar{R}^0_{\text{UComp}}(n).$